\documentclass[useAMS,usenatbib]{mn2e}
\usepackage{subfigure}
\usepackage{graphicx, amssymb}
\usepackage[fleqn]{amsmath}
\usepackage{epsfig}
\usepackage{color}
\usepackage{times}
\usepackage{natbib}

\title[Redshift evolution of extragalactic rotation measures]
{Redshift evolution of extragalactic rotation measures}
\author[J. Xu and J. L. Han]{J. Xu$^{1,2}$ and J. L. Han$^{1,3}$\thanks{E-mail: hjl@nao.cas.cn}\\
$^{1}$National Astronomical Observatories, Chinese Academy of Sciences,
       A20 Datun Road, Chaoyang District, Beijing 100012, China.\\
$^{2}$School of Physics, University of Chinese Academy of Sciences, 
       Beijing 100049, China.\\
$^{3}$School of Astronomy and Space Science, Nanjing University, Nanjing, China}
\begin{document}

\date{Accepted 2014 May 19. Received 2014 May 19; in original form 2014 March 19
}

\pagerange{\pageref{firstpage}--\pageref{lastpage}} \pubyear{?}

\maketitle

\label{firstpage}

\begin{abstract}
We obtained rotation measures of 2642 quasars by cross-identification
of the most updated quasar catalog and rotation measure catalog. After
discounting the foreground Galactic Faraday rotation of the Milky Way,
we get the residual rotation measure (RRM) of these quasars. We
carefully discarded the effects from measurement and systematical
uncertainties of RRMs as well as large RRMs from outliers, and get
marginal evidence for the redshift evolution of real dispersion of
RRMs which steady increases to 10 rad~m$^{-2}$ from $z=0$ to $z\sim1$
and is saturated around the value at higher redshifts. The ionized
clouds in the form of galaxy, galaxy clusters or cosmological
filaments could produce the observed RRM evolutions with different
dispersion width. However current data sets can not constrain the
contributions from galaxy halos and cosmic webs. Future RM
measurements for a large sample of quasars with high precision are
desired to disentangle these different contributions.
\end{abstract}

\begin{keywords}
polarization --- intergalactic medium --- radio continuum: general --- 
magnetic fields
\end{keywords}

\section{Introduction}

Faraday rotation is a powerful tool to probe the extragalactic medium.
The observed rotation measure of a linearly polarized radio source at
redshift $z_{\rm s}$ is determined by the polarization angle rotation
($\psi_1 - \psi_2$) against the wavelength square ($\lambda_1^2-
\lambda_2^2$)
\begin{equation}
RM_{\rm obs} = \frac{\psi_1 - \psi_2}{\lambda_1^2- \lambda_2^2}
= 0.81 \int_{z_s}^0\frac{n_e(z)B_{||}(z)}{(1+z)^{2}}\frac{dl}{dz} dz.
\label{angle}
\end{equation}
The rotation measure (RM, in the unit of rad m$^{-2}$) is an
integrated quantity of the product of thermal electron density ($n_{\rm
  e}$, in the unit of cm$^{-3}$) and magnetic fields along the line of
sight ($B_{||}$, in the unit of $\mu$G) over the path from the source
at a redshift $z_{\rm s}$ to us. Here the comoving path increment per
unit redshift, $dl/dz$, is in parsecs.
The observed rotation measure, $RM_{\rm obs}$, with a uncertainty,
$\sigma_{RM}$, is a sum of the rotation measure intrinsic to the
source, $RM_{\rm in}$, the rotation measure in intergalactic space,
$RM_{\rm IGM}$, the foreground Galactic RM, $GRM$, from our Milky Way
Galaxy, i.e.
\begin{equation}
RM_{\rm obs}=RM_{\rm in}+RM_{\rm IGM} + GRM.
\label{rmobs}
\end{equation}
It has been found that the RM distribution of radio sources in the sky
are correlated in angular scale of a few degree to a few tens degree
\citep{sk80,ow95,hmbb97,sts11}, which indicates the smooth Galactic RM
foreground. The extragalactic rotation measures is $RM_{\rm
  in}+RM_{\rm IGM} = RM_{\rm obs}- GRM$, which is often called as
residual rotation measure (RRM), i.e. the residual after the
foreground Galactic RM is discounted from the observed RM. Because the
polarization angle undergoes a random walk in the intergalactic space
due to intervening magnetoionic medium, the RRMs from the
intergalactic medium should have a zero-mean Gaussian distribution.
Radio sources at higher redshift will pass through more intervening
medium, so that variance of RRMs, $V_{\rm RRM}$, of a sample of
sources is expected to get larger at higher redshifts. Though the
measured RM values from a source could be likely wavelength dependent
due to unresolved multiple components \citep{frb11,xh12,bml12}, RM
values intrinsic to a radio source at redshift $z_{\rm s}$ are reduced
by a factor $(1+z_{\rm s})^{−2}$ due to change of $\lambda$ when
transformed to the observer's frame, and for the variance by a factor
$(1+z_{\rm s})^{−4}$, the RRMs are therefore often statistically used
to probe magnetic fields in the intervening medium between the source
and us, such as galaxies, galaxy clusters or cosmic webs.
\begin{figure}
\includegraphics[angle=-90,width=82mm]{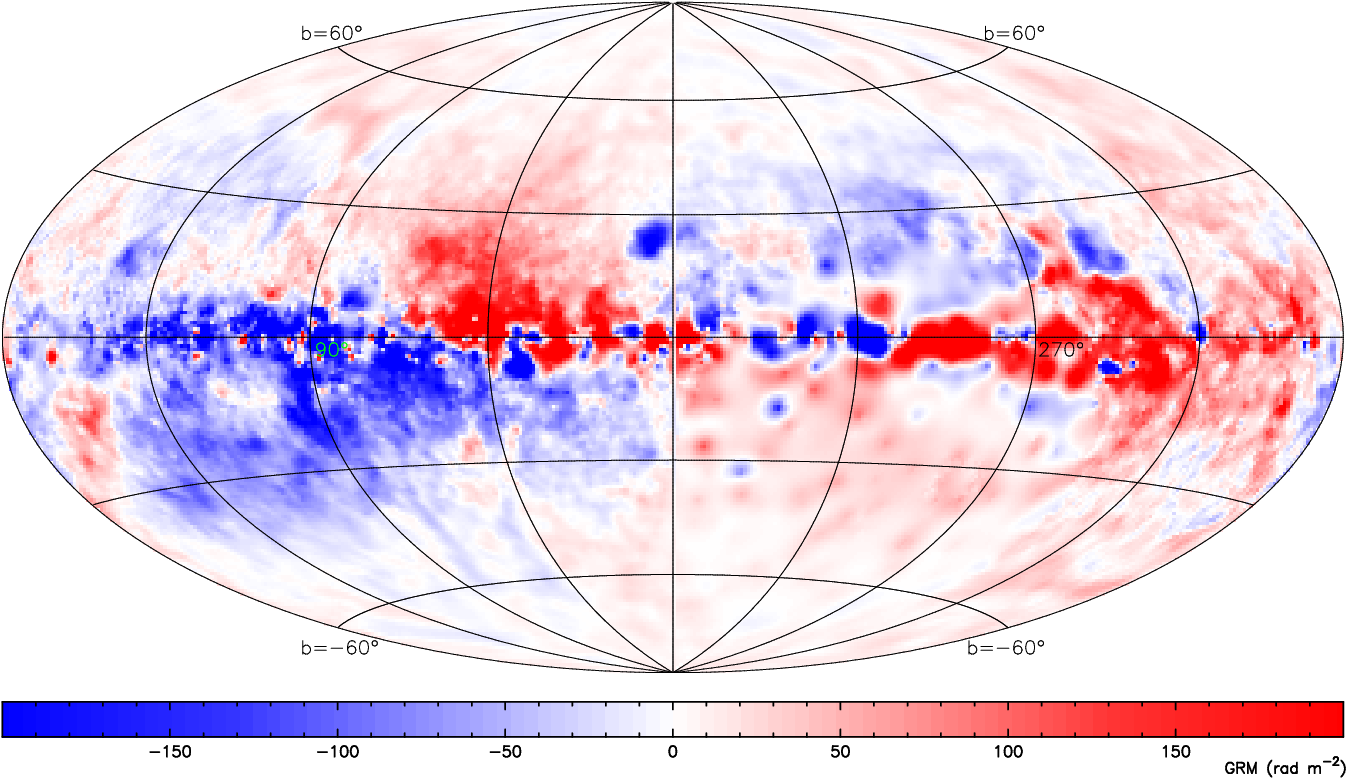}\\ 
\includegraphics[angle=-90,width=82mm]{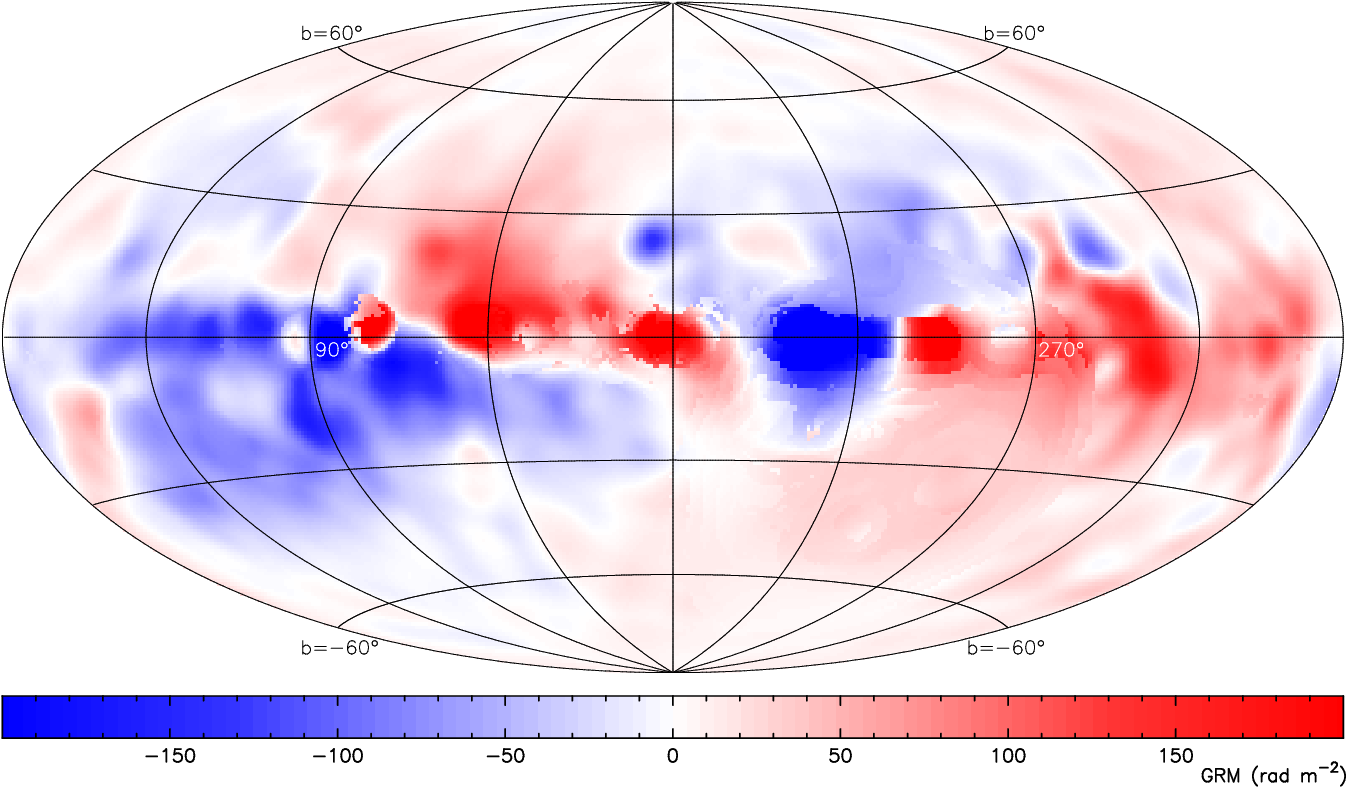} \\
\includegraphics[angle=-90,width=82mm]{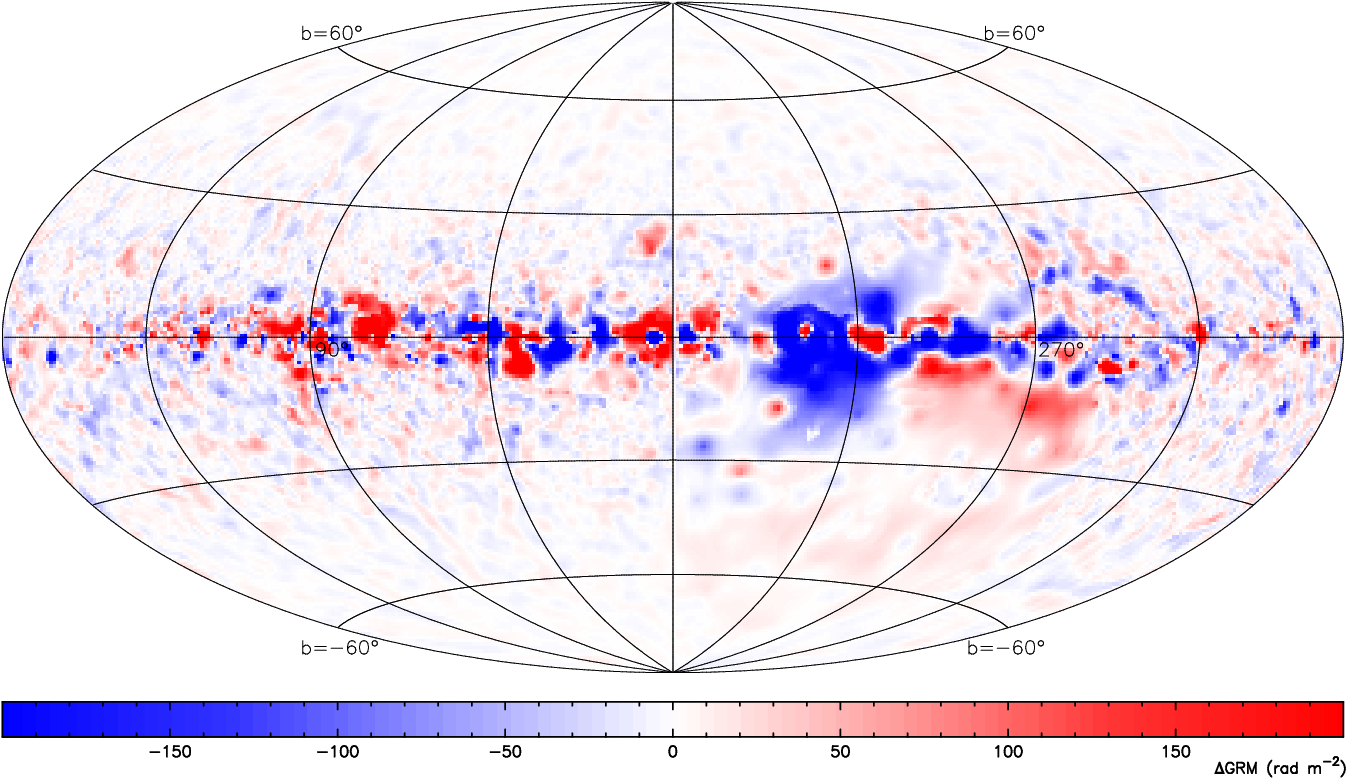}\\
\caption{Sky view for the foreground Galactic RM derived by
  \citet{ojr+12} (the top panel) and by \citet{xh14} (the middle
  panel), and their difference (the bottom panel). See \citet{xh14}
  for details.}
\label{fig1}
\end{figure}

Previously there have been many efforts to investigate RRM
distributions and their possible evolution with redshift. Without a
good assessment of the foreground Galactic RM in early days, RMs of a
small sample of radio sources gave some indications for larger RRM
data scatter at higher redshifts, which were taken as evidence of
magnetic field in the intergalactic medium
\citep{nel73,val75,ks76,krs77,tn82}. \citet{bur74} proposed the
steady-state model and found that the variance of RRM approaches a
limiting value at $z_s\sim 1$; \citet{krs77} suggested $\sigma_{\rm
  IGM}<10$ rad~m$^{-2}$; \citet{val75} claimed the upper limit of
intergalactic rotation measure as being 10 rad~m$^{-2}$. Theoretical
models for the random intergalactic magnetic fields in the Friedmann
cosmology \citep{nel73} and in Einstein-de Sitter cosmology
\citep{bur74} and for the uniform fields \citep{val75} have been
proposed. \citet{tn82} summarized the Friedmann model \citep{nel73}
and the steady-state model \citep{bur74} and also proposed their own
ionized cloud model. In Friedmann model, particles conservation is
assumed and the field is frozen in the evolving Friedmann cosmology,
\citet{tn82} showed $V_{\rm RRM}$ increasing with $(1+z)^{3 \sim 4}$
depending on cosmology density $\Omega_M$. In the steady-state model
contiguous random cells do not vary with time, which induces the
intergalactic $V_{\rm RRM} \propto [1-(1+z_s)^{-4}]$. In the ionized
cloud model, the Faraday-active cells with random fields are in the
form of non-evolving discrete gravitationally bound, ionized clouds,
so that the final $V_{\rm RRM} \propto [1-(1+z_s)^{-2}]$. \citet{tn82}
applied these three models to fit the increasing RM variance of 134
quasars against redshift, but can not distinguish the models due to
large uncertainties. \citet{wpk84} made a statistics of the RRMs of
112 quasars and found a systematic increase of $V_{\rm RRM}$ with
redshift even up to redshift $z$ above 2. After considering possible
contributions to the RRM variance from RMs intrinsic to quasars or
from RMs due to discrete intervening clouds, \citet{wpk84} suggest
that the observed RRM variance mainly results from absorption-line
associated intervening clouds.

Because intervening galaxies are most probable clouds for the
intergalactic RMs at cosmological distances, efforts to search for
evidence for the association between the enlarged RRM variance with
optical absorption-lines of quasars therefore have been made for many
years, first by \citet{kp82}, later by
\citet{wpk84,wp91,wlo92,ow95,bml+08,kbm+08,bml10}, and most recently
by \citet{bml12} and \citet{jc13}. Small and later larger quasar
samples with or without the MgII absorption lines
\citep[e.g.][]{jc13}, with stronger or weaker MgII absorption lines
\citep[e.g.][]{bml10}, with or without the Ly$\alpha$ absorption lines
\citep[e.g.][]{ow95}, are compared for RRM distributions. In almost
all cases, the RRM or absolute values of RMs of quasars with
absorption lines show significantly different cumulative RM
probability distribution function or a different variance value from
those without absorption lines, and those of higher redshift quasars
show a marginally significant excess compared to that of lower redshift
objects.  Most recently \citet{jc13} got the excess of RRM deviation
of $8.1\pm4.8$~rad~m$^{-2}$ for quasars with MgII absorption-lines.

Certainly intervening objects could be large-scale cosmic-web or
filaments or super-clusters of galaxies, with a coherence length much
larger than a galaxy, which may result in a possible excess of RRMs
\citep{xkh+06}. At least the RRM excess due to galaxy clusters has
been statistically detected \citep[e.g.][]{ckb01,gdm+10}. Computer
simulations for large-scale turbulent magnetic fields together with
inhomogeneous density in the cosmic web of tens of Mpc scale have been
tried by, e.g., \citet{bbo99,rkcd08,ar10,ar11}, and also compared
with real RRM data. The RMs from cosmic web probably are very small,
only about a few rad~m$^{-2}$ \citep{ar11}. The dispersion of
so-caused RRMs is also small, which increases steeply for $z < 1$ and
saturates at a value of a few rad~m$^{-2}$ at $z \sim 1$.

Because of the smallness of RM contribution from intergalactic space,
to study the redshift evolution of extragalactic RMs, we have to
enlarge the sample size of high redshift objects for RMs and have to
reduce the RRM uncertainty. The uncertainty of RRM is limited by not
only the observed accuracy for RMs of radio sources but also the
accuracy of estimated foreground of the Galactic RMs. The RMs were
found to be correlated over a few tens of degree in the mid-latitude
area \citep[e.g.][]{sk80,ow95}. The GRM uncertainty in most previous
studies is large, around 20 rad~m$^{-2}$ in general, due to a small
covering density of available RMs in the sky. Noticed that RMs have
smallest random values near the two Galactic poles
\citep{sk80,hmq99,mgh+10}. To reduce the uncertainty of RRMs,
\citet{yhc03} tried to use RMs of only 43 carefully selected
extragalactic radio sources toward Galactic poles, and found only the
marginal increase of $V_{\rm RRM}$ with redshift.

\begin{table*}
\centering
\caption{2642 quasars with RM data available in literature}
\label{qsosample}
\begin{tabular}{rrrrrrrrrrrr} 
\hline
     RA~~  &  Dec~~    & z~~~    & GL~~     & GB~~      &  RM~~        &$\sigma_{RM}$~~ & Ref~~ &GRM~~       &$\sigma_{GRM}$~~ & RRM~~       &$\sigma_{RRM}$~~\\
    (deg)  &  (deg)    &         &   (deg) &  (deg)    &(rad m$^{-2}$) &(rad m$^{-2}$) &     &(rad m$^{-2}$)&(rad m$^{-2}$)   &(rad m$^{-2}$)&(rad m$^{-2}$)   \\ 
\hline
    0.0417 &   30.9331 &  1.801  & 110.1507 & --30.6630  &   --37.9     &  11.0  &  tss09 &   --68.8   &   1.7  &   30.9  &   11.1\\
    0.2542 &   24.1450 &  0.300  & 108.4335 & --37.3031  &   --63.2     &  14.5  &  tss09 &   --57.8   &   1.9  &   --5.4  &   14.6\\
    0.3867 &   14.9356 &  0.399  & 105.3749 & --46.2285  &   --34.9     &   3.8  &  tss09 &   --17.0   &   1.2  &  --17.9  &    4.0\\
    0.7050 &   30.5447 &  2.300  & 110.6968 & --31.1693  &   --40.5     &  13.9  &  tss09 &   --68.6   &   1.7  &   28.1  &   14.0\\
    0.7992 &   16.4839 &  1.600  & 106.5177 & --44.8449  &   --24.3     &   8.5  &  tss09 &   --21.2   &   1.2  &   --3.1  &    8.6\\
    0.9283 &  --11.8633 &  1.300  &  84.3539 & --71.0677  &    --3.8     &  13.3  &  tss09 &     0.4   &   1.4  &   --4.2  &   13.4\\
    0.9383 &  --11.1383 &  1.569  &  85.6081 & --70.4718  &    --8.2     &   9.3  &  tss09 &     1.8   &   1.4  &  --10.0  &    9.4\\
    1.3108 &    4.4186 &  1.200  & 101.7086 & --56.5377  &    13.3     &   5.6  &  tss09 &    --2.7   &   1.4  &   16.0  &    5.8\\
    1.5942 &   --0.0733 &  1.038  &  99.2808 & --60.8590  &    12.0     &   3.0  &  skb81 &    --5.2   &   1.6  &   17.2  &    3.4\\
    1.5958 &   12.5981 &  0.980  & 106.1113 & --48.7967  &   --11.2     &   8.5  &  tss09 &   --10.0   &   1.2  &   --1.2  &    8.6\\
    1.6471 &    8.8044 &  1.900  & 104.5495 & --52.4592  &    --8.0     &  13.7  &  tss09 &    --3.5   &   1.2  &   --4.5  &   13.7\\
    2.0550 &   13.6133 &  1.000  & 107.1538 & --47.9300  &     0.1     &  15.9  &  tss09 &   --11.9   &   1.2  &   12.0  &   15.9\\
    2.1925 &    0.0611 &  0.505  & 100.5304 & --60.9412  &   --38.5     &  18.8  &  tss09 &    --5.4   &   1.5  &  --33.1  &   18.9\\
    2.2071 &   --0.2778 &  2.000  & 100.3199 & --61.2648  &     7.8     &   8.0  &  tss09 &    --5.3   &   1.5  &   13.1  &    8.1\\
    2.2662 &    6.4725 &  0.400  & 104.4242 & --54.8694  &   --17.1     &   7.4  &  tss09 &    --1.8   &   1.3  &  --15.3  &    7.5\\
    2.4463 &    6.0972 &  2.311  & 104.5382 & --55.2800  &    10.3     &  15.4  &  tss09 &    --1.7   &   1.3  &   12.0  &   15.5\\
    2.5758 &   14.5606 &  0.901  & 108.2184 & --47.1326  &   --25.6     &  10.5  &  tss09 &   --14.1   &   1.1  &  --11.5  &   10.6\\
    2.6196 &   20.7969 &  0.600  & 110.1993 & --41.0599  &   --36.4     &  16.2  &  tss09 &   --36.6   &   1.6  &    0.2  &   16.3\\
    2.6450 &  --30.9042 &  0.999  &   7.5916 & --80.3078  &   --10.4     &   9.3  &  tss09 &     8.3   &   0.8  &  --18.7  &    9.3\\
    2.8967 &    8.3986 &  1.300  & 106.3363 & --53.1842  &     3.2     &   2.0  &  tss09 &    --2.9   &   1.2  &    6.1  &    2.3\\
    3.0346 &    7.3308 &  1.800  & 106.0930 & --54.2510  &   --22.1     &  14.5  &  tss09 &    --2.1   &   1.2  &  --20.0  &   14.6\\
    3.3363 &  --15.2297 &  1.838  &  84.3517 & --75.1678  &    11.1     &  13.0  &  tss09 &    --0.1   &   1.2  &   11.2  &   13.1\\
    3.4754 &   --4.3978 &  1.075  &  99.7867 & --65.5687  &    --1.5     &   4.4  &  tss09 &    --0.0   &   1.4  &   --1.5  &    4.6\\
    3.6575 &  --30.9886 &  2.785  &   5.1071 & --81.0824  &     9.0     &   2.0  & mgh+10 &     7.7   &   0.8  &    1.3  &    2.2\\
    3.7604 &  --18.2142 &  0.743  &  77.7952 & --77.7642  &    --2.5     &   4.9  &  tss09 &     3.7   &   1.0  &   --6.2  &    5.0\\
    4.0000 &   39.0072 &  1.721  & 115.4420 & --23.3486  &  --123.9     &   4.0  & kmg+03 &   --81.6   &   2.4  &  --42.3  &    4.7\\
    4.0533 &   29.7517 &  1.300  & 113.8648 & --32.4992  &   --74.3     &   8.1  &  tss09 &   --66.0   &   1.7  &   --8.3  &    8.3\\
    4.0729 &   24.9656 &  1.800  & 112.9169 & --37.2218  &   --43.1     &  13.6  &  tss09 &   --60.1   &   1.8  &   17.0  &   13.7\\
    4.1658 &   25.1747 &  1.300  & 113.0655 & --37.0299  &   --77.1     &   6.9  &  tss09 &   --60.9   &   1.8  &  --16.2  &    7.1\\
    4.2588 &   32.1558 &  1.086  & 114.5124 & --30.1529  &   --42.1     &  12.5  &  tss09 &   --62.4   &   1.6  &   20.3  &   12.6\\
\hline
\end{tabular}
\\This table is available in its entirety online. A portion is shown here for guidance regarding its form and content.
\end{table*}

In addition to the previously cataloged RMs
\citep[e.g. ][]{skb81,bmv88} and published RM data in literature,
\citet{tss09} have reprocessed the 2-band polarization data of the
NRAO VLA Sky Survey \citep[NVSS,][]{ccg+98}, and obtained the two-band
RMs for 37,543 sources. Though there is a systematical uncertainty of
$10.0\pm1.5$ rad~m$^{-2}$ \citep{xh14}, the NVSS RMs can be used
together to derive the foreground Galactic RM \citep{ojr+12,xh14}, see
Fig.~\ref{fig1}. \citet{hrg12} obtained the RMs of a sample of 4003
extragalactic objects with known redshifts (including 860 quasars,
data not released yet) by cross-identification of the NVSS RM catalog
sources \citep{tss09} with known optical counterparts (galaxies, AGNs
and quasars) in literature, and they concluded that the variance for
RRMs does not evolve with redshift. Nevertheless, \citet{nsb13} used
the same dataset and found strong evidence for the redshift evolution
of absolute values of RMs. Further investigation on this controversy
is necessary.

Recently, \citet{xh14} compiled a catalog of reliable RMs for 4553
extragalactic point radio sources, and used a weighted average method
to calculate the Galactic RM foreground based on the compiled RM data
together with the NVSS RM data. On the other hand, a new version of
quasar catalog (Milliquas) is updated and available on the
website\footnote{http://quasars.org/milliquas.htm}, which compiled
about 1,252,004 objects from literature and archival surveys and
databases. Here we cross-identify the two large datasets, and obtained
a large sample of RMs for 2642 quasars, which can be used to study the
redshift evolution of extragalatic RMs. We will introduce data in the
Section 2, and study their distribution in Section 3. We discuss the
results and fit the models in Section 4.

\section{Rotation measure data of quasars}

We obtained the rotation measure data of quasars from the
cross-identification of quasars in the newest version of the Million
Quasars (Milliquas) catalog with radio sources in the NVSS RM catalog
\citep{tss09} and the compiled RM catalog \citep{xh14}. The Million
Quasars catalog (version 3.8a, Eric Flesch, 2014) is a compilation of
all known type~I quasars, AGN, and BL-Lacs in literature. To avoid
possible influence on RRMs from different polarization fractions of
galaxies and quasars \citep{hrg12}, we take only type~I quasars in the
catalog. We adopt 3$''$ as the upper limit of position offset for
associations between quasars and the radio sources with rotation
measure data, according to \citet{hrg12}, and finaly get RMs for 2642
associated quasars, as listed in Table~\ref{qsosample}, which is the
largest dataset of quasar RMs up to date.

\begin{figure*}
\centering
\includegraphics[angle=-90,width=80mm]{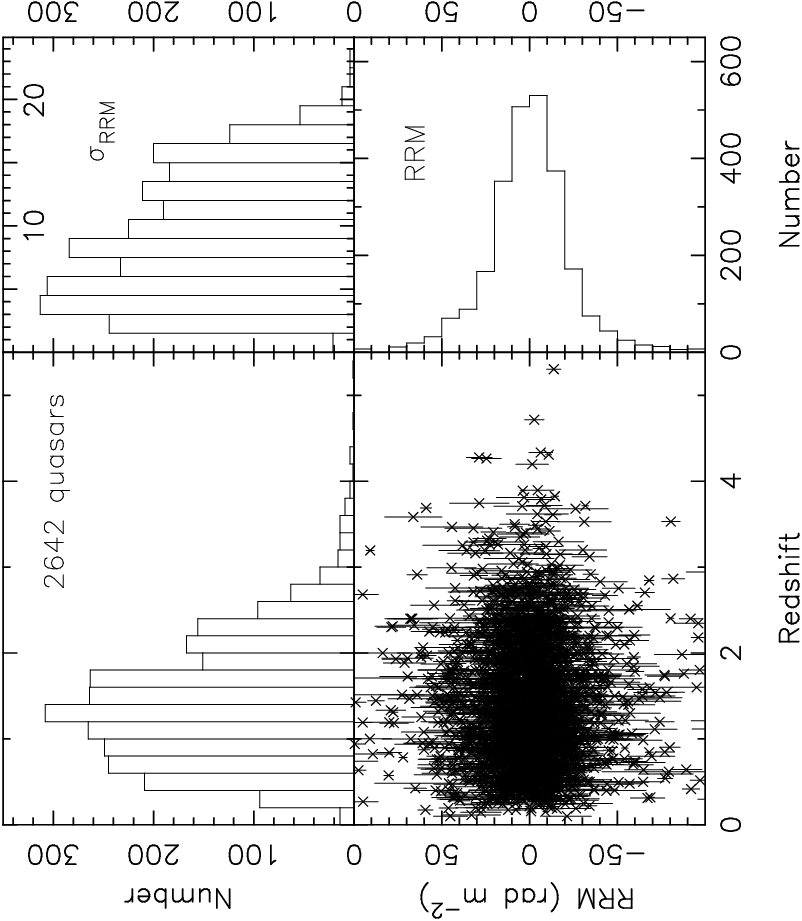} \hspace{1mm}
\includegraphics[angle=-90,width=80mm]{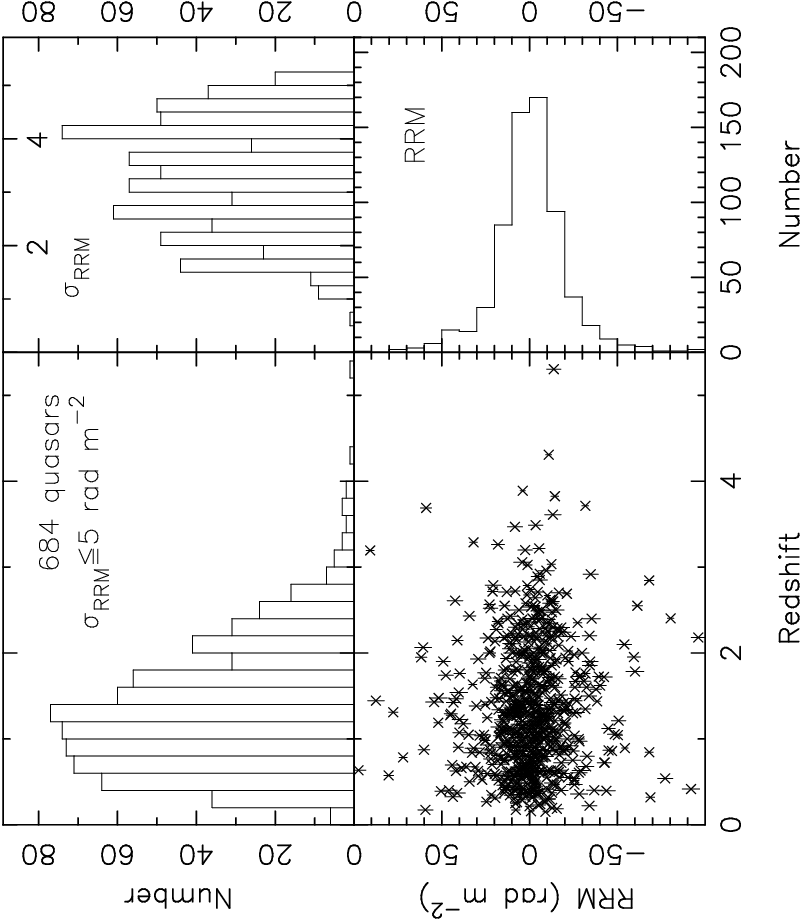} \\
\includegraphics[angle=-90,width=80mm]{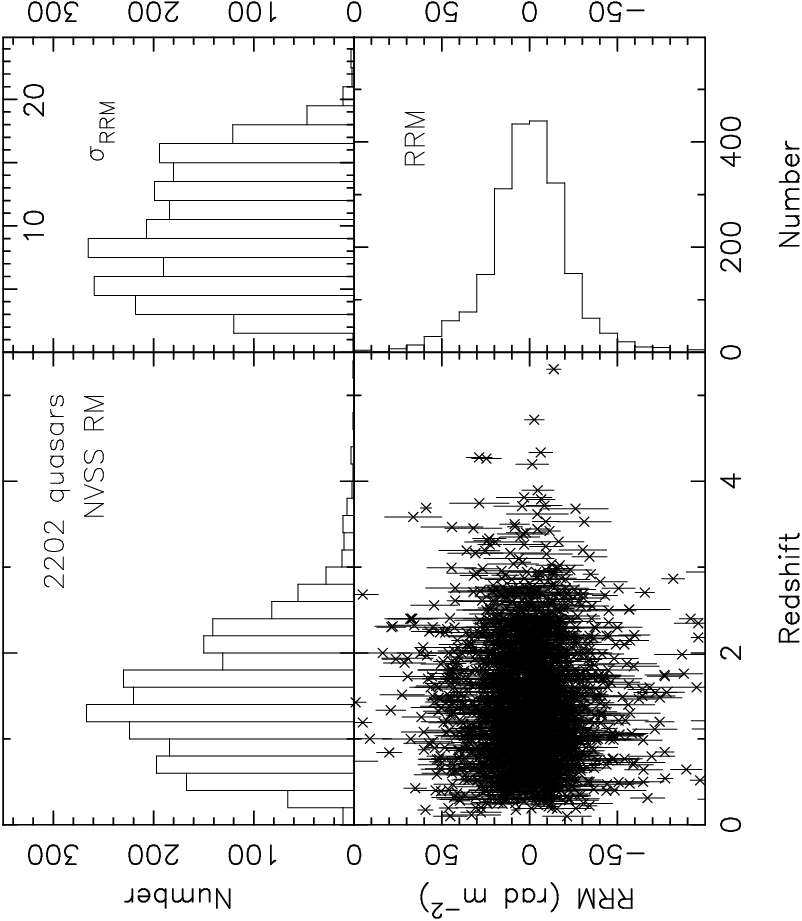}  \hspace{1mm}
\includegraphics[angle=-90,width=80mm]{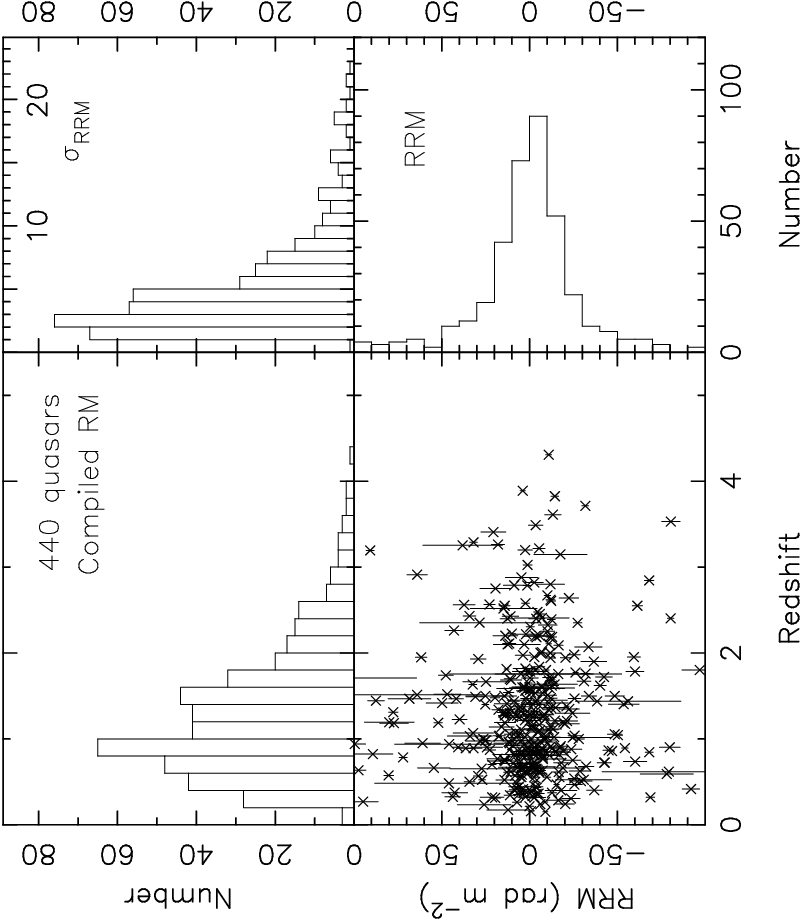} 
\caption{The redshift and RRM distribution for 2642 quasars together
  with the histogram for RRM uncertainty ({\it upper left panel}) and
  684 quasars with formal RM uncertainty $\sigma_{RRM} \leqslant$ 5
  rad~m$^{-2}$ ({\it upper right panel}).  Similar plots for 2202
  quasars with only the NVSS RMs of \citet{tss09} ({\it lower left
    panel}) or for 440 quasars with RMs from the compiled RM catalog
  of \citet{xh14} ({\it lower right panel}). }
\label{fig2}
\end{figure*}

To get the extragalactic rotation measures of these quasars, we have
to discount the foreground RM from our Milky Way. The foreground
Galactic RMs (GRM) vary with the Galactic longitude (GL) and latitude
(GB). In recent similar works \citep[e.g.][]{hrg12,nsb13} the
foreground RMs were taken from estimations of \citet{ojr+12} using a
complicated signal reconstruction algorithm within the framework of
the information field theory. We have used an improved weighted
average method to estimate the Galactic RM foreground by using the
cleaned RM data without outliers, which gives more reliable
estimations of the GRM with smaller uncertainties \citep{xh14}. The
extragalactic rotation measures, i.e the residual rotation measures,
is then obtained by $RRM = RM-GRM$, and their uncertainty by
$\sigma_{\rm RRM}=\sqrt{\sigma_{\rm RM}^2+\sigma_{\rm GRM}^2}$, as
listed in Table~\ref{qsosample}.

The RRM distributions of 2642 quasars are shown in Figure \ref{fig2},
including the distribution against redshift and amplitude, together
with the histograms for RRM amplitude and uncertainty. Most RM data
taken from \citet{tss09} have a large formal uncertainty and also a
previously unknown systematic uncertainty \citep{xh14,mgh+10}. Because
the uncertainty is a very important factor for deriving the redshift
evolution of the residual rotation measures (see below), the best RRM
data-set for redshift evolution study should be these with a very
small uncertainty, e.g. $\sigma_{\rm RRM} \leqslant$ 5
rad~m$^{-2}$. We get a RRM data-set of 684 quasars with such a formal
accuracy, without considering the systematic uncertainty, and their
RRM distribution is shown in the upper right panel in
Figure~\ref{fig2}. To clarify the sources of RM data, we present the
RRM distribution for 2202 quasars which have the RM values obtained
only from the NVSS RM catalog \citep{tss09}, and also for 440 quasars
whose RM values are obtained from the compiled RM catalog of
\citet{xh14}.

\begin{table*}
\centering
\caption{Statistics for real RRM distribution of subsamples in redshift bins}
\label{dataresult}
\begin{tabular}{c|rccr|rccr} 
\hline
\multicolumn{1}{c|}{ } & \multicolumn{4}{c|}{Subsamples from the NVSS RM catalog}       &     \multicolumn{4}{c}{Subsamples from the compiled RM catalog}         \\         
\hline
Redshift &  No. of   & z$_{median}$   & $W_{\rm RRM}$ & $W_{\rm RRM0}$~~~&  
            No. of   & z$_{median}$   & $W_{\rm RRM}$ & $W_{\rm RRM0}$~~~~~~\\          
range    &  quasars  &               & (rad~m$^{-2}$) & (rad~m$^{-2}$)  &
            quasars  &   & (rad~m$^{-2}$)  & (rad~m$^{-2}$)   \\
\hline                                                            
\multicolumn{1}{c|}{ } & \multicolumn{8}{c}{2338 quasars of $\sigma_{\rm RRM} \leqslant$ 20 rad~m$^{-2}$: 2018 NVSS RMs and 320 compiled RMs}\\
\hline
0.0--0.5 &   152     & 0.400         &   17.7$\pm$2.9 &14.6$\pm$2.9       &      38     & 0.356         &   11.2$\pm$4.5 &10.8$\pm$4.5              \\
0.5--1.0 &   455     & 0.772         &   17.5$\pm$1.8 &14.4$\pm$1.8       &     109     & 0.768         &   13.2$\pm$5.6 &12.9$\pm$5.6              \\
1.0--1.5 &   587     & 1.286         &   17.4$\pm$1.8 &14.2$\pm$1.8       &      82     & 1.276         &   15.0$\pm$3.1 &14.7$\pm$3.1              \\
1.5--2.0 &   441     & 1.756         &   18.2$\pm$1.8 &15.2$\pm$1.8       &      47     & 1.685         &   14.5$\pm$8.6 &14.2$\pm$8.6              \\
2.0--3.0 &   383     & 2.300         &   17.5$\pm$1.9 &14.4$\pm$1.9       &      44     & 2.396         &   14.3$\pm$6.9 &14.0$\pm$6.9              \\
\hline 
\multicolumn{1}{c|}{ } & \multicolumn{8}{c}{2015 quasars of $\sigma_{\rm RRM} \leqslant$ 15 rad~m$^{-2}$: 1703 NVSS RMs and 312 compiled RMs}\\
\hline
0.0--0.5 &   136     & 0.400         &   16.9$\pm$3.4 &13.6$\pm$3.4       &      37     & 0.360         &   10.7$\pm$4.4 &10.3$\pm$4.4              \\
0.5--1.0 &   386     & 0.768         &   17.1$\pm$2.0 &13.9$\pm$2.0       &     106     & 0.775         &   13.1$\pm$5.6 &12.8$\pm$5.6              \\
1.0--1.5 &   510     & 1.286         &   17.0$\pm$1.8 &13.7$\pm$1.8       &      81     & 1.271         &   15.2$\pm$3.1 &14.9$\pm$3.1              \\
1.5--2.0 &   365     & 1.741         &   18.0$\pm$2.0 &15.0$\pm$2.0       &      46     & 1.690         &   14.7$\pm$9.1 &14.4$\pm$9.1              \\
2.0--3.0 &   306     & 2.300         &   17.4$\pm$2.5 &14.2$\pm$2.5       &      42     & 2.371         &   14.4$\pm$7.2 &14.1$\pm$7.2              \\
\hline  

\multicolumn{1}{c|}{ } & \multicolumn{8}{c}{1425 quasars of  $\sigma_{\rm RRM} \leqslant$ 10 rad~m$^{-2}$: 1129 NVSS RMs and 296 compiled RMs}\\
\hline 
0.0--0.5 &    88     & 0.400         &   15.4$\pm$3.4 &11.7$\pm$3.4       &      36     & 0.362         &   10.8$\pm$4.4 &10.4$\pm$4.4              \\
0.5--1.0 &   272     & 0.752         &   16.4$\pm$2.1 &13.0$\pm$2.1       &      99     & 0.799         &   13.3$\pm$6.2 &13.0$\pm$6.2              \\
1.0--1.5 &   336     & 1.283         &   15.7$\pm$2.0 &12.1$\pm$2.0       &      76     & 1.270         &   14.8$\pm$3.0 &14.5$\pm$3.0              \\
1.5--2.0 &   232     & 1.724         &   17.5$\pm$2.1 &14.4$\pm$2.1       &      43     & 1.700         &   13.9$\pm$9.9 &13.6$\pm$9.9              \\
2.0--3.0 &   201     & 2.300         &   16.6$\pm$2.6 &13.2$\pm$2.6       &      42     & 2.371         &   14.4$\pm$7.2 &14.1$\pm$7.2              \\
\hline 
\multicolumn{1}{c|}{ } & \multicolumn{8}{c}{626 quasars of $\sigma_{\rm RRM} \leqslant$ 5 rad~m$^{-2}$: 406 NVSS RMs and 220 compiled RMs}\\
\hline    
0.0--0.5 &    40     & 0.394         &   13.7$\pm$4.2 & 9.4$\pm$4.2       &      27     & 0.364         &    8.4$\pm$ 2.7& 7.8$\pm$ 2.7             \\
0.5--1.0 &    93     & 0.720         &   15.1$\pm$4.7 &11.3$\pm$4.7       &      77     & 0.751         &   12.7$\pm$ 6.1&12.3$\pm$ 6.1             \\
1.0--1.5 &   119     & 1.270         &   13.6$\pm$3.1 & 9.2$\pm$3.1       &      56     & 1.268         &   13.6$\pm$ 3.0&13.3$\pm$ 3.0             \\
1.5--2.0 &    76     & 1.732         &   15.1$\pm$4.4 &11.3$\pm$4.4       &      34     & 1.704         &   14.4$\pm$12.4&14.1$\pm$12.4             \\
2.0--3.0 &    78     & 2.316         &   13.4$\pm$3.6 & 8.9$\pm$3.6       &      26     & 2.396         &   13.9$\pm$14.8&13.6$\pm$14.8             \\
\hline
\end{tabular}
\end{table*}

\begin{figure*}
\centering
\includegraphics[angle=-90,width=33mm]{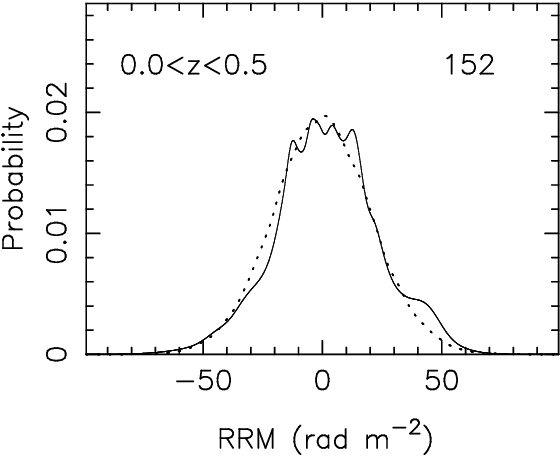} 
\includegraphics[angle=-90,width=33mm]{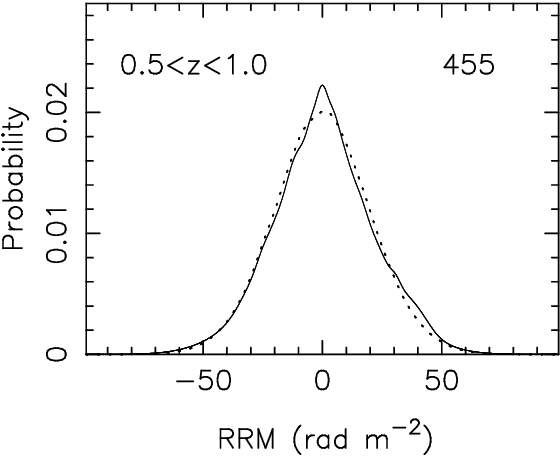} 
\includegraphics[angle=-90,width=33mm]{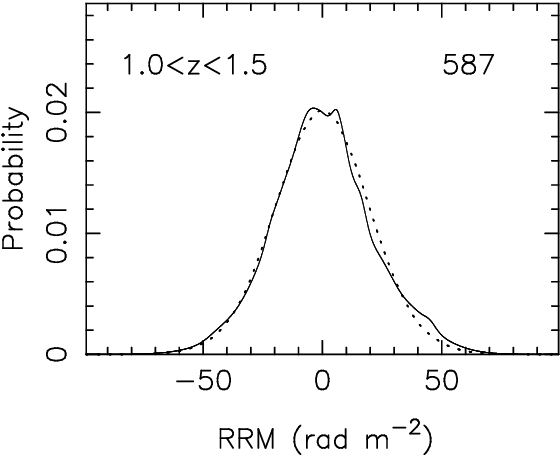} 
\includegraphics[angle=-90,width=33mm]{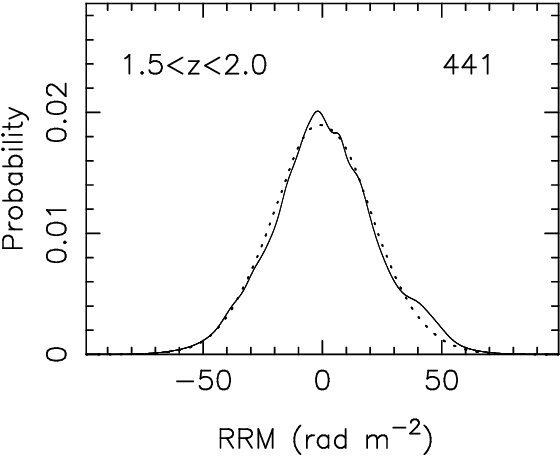} 
\includegraphics[angle=-90,width=33mm]{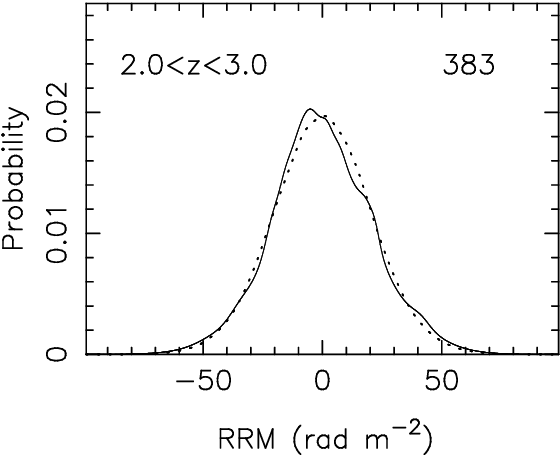}
\includegraphics[angle=-90,width=33mm]{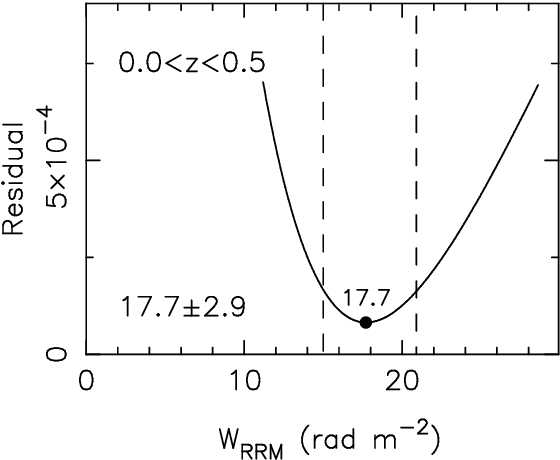}  
\includegraphics[angle=-90,width=33mm]{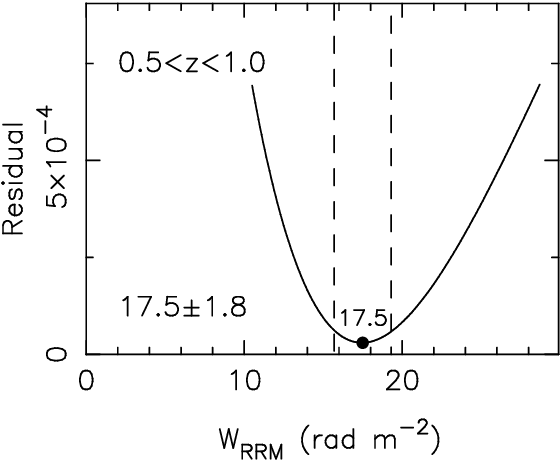}  
\includegraphics[angle=-90,width=33mm]{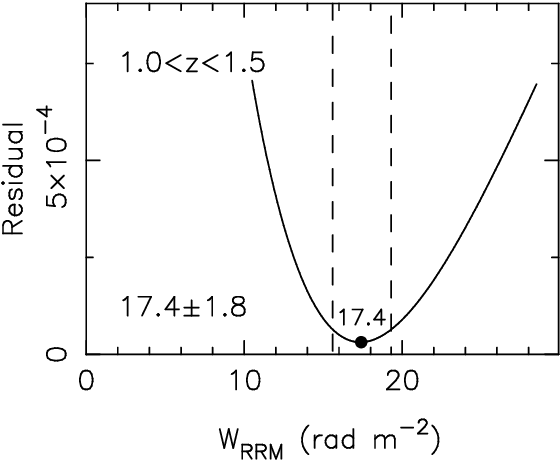}  
\includegraphics[angle=-90,width=33mm]{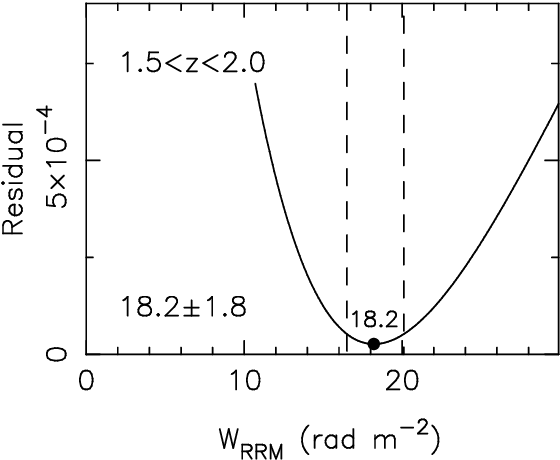}  
\includegraphics[angle=-90,width=33mm]{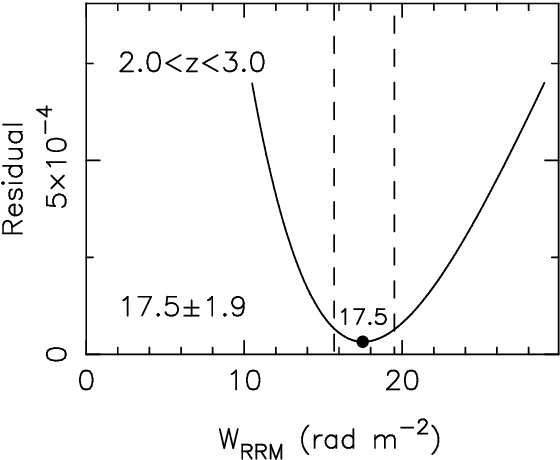} 
\caption{The probability distribution function of measured RRM values
  [solid line for $P(RRM)$] compared with that of the mock RRM sample
  with the best $W_{\rm RRM}$ [dotted line for $P_{\rm mock}(RRM)$]
  for the subsamples of quasars with the NVSS RMs and $\sigma_{\rm
    RRM}\le 20$~rad~m$^{-2}$ in different redshift ranges. The fitting
  residues, which is mimic to $\chi^2$, against various $W_{\rm RRM}$
  are plotted in the lower panels, which define the best $W_{\rm RRM}$
  and its uncertainty at 68\% probability.}
\label{MockRRM20}
\end{figure*}

\begin{figure*}
\centering
\includegraphics[angle=-90,width=33mm]{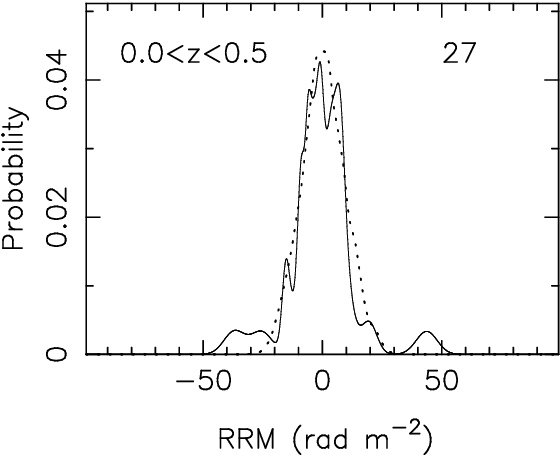} 
\includegraphics[angle=-90,width=33mm]{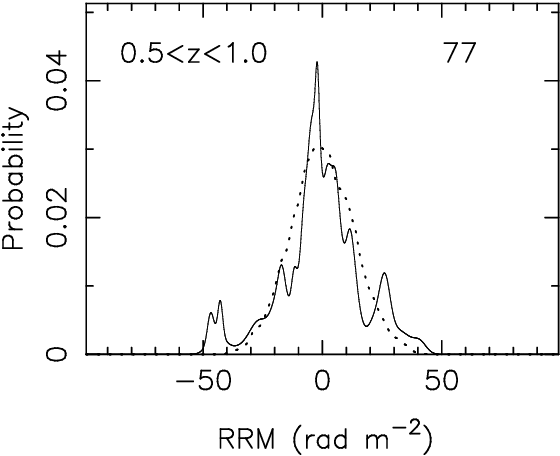} 
\includegraphics[angle=-90,width=33mm]{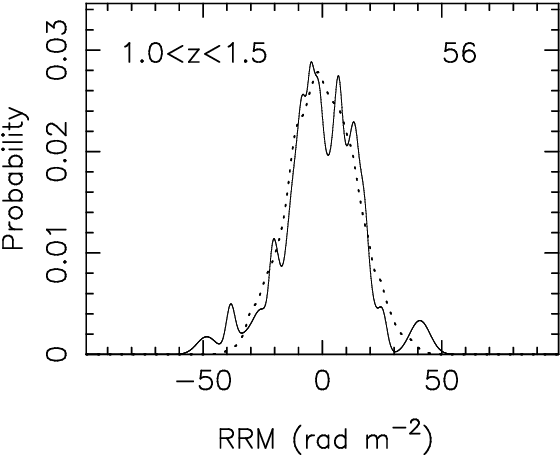} 
\includegraphics[angle=-90,width=33mm]{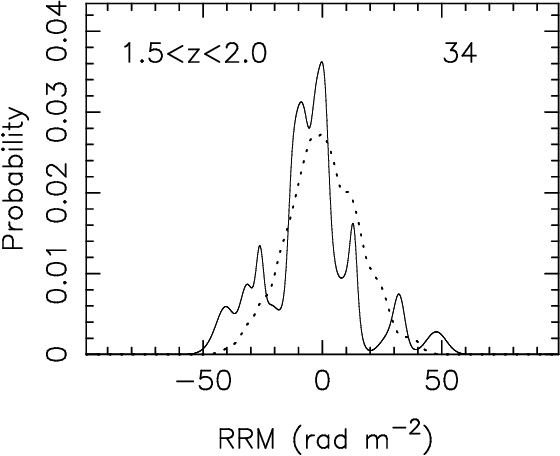} 
\includegraphics[angle=-90,width=33mm]{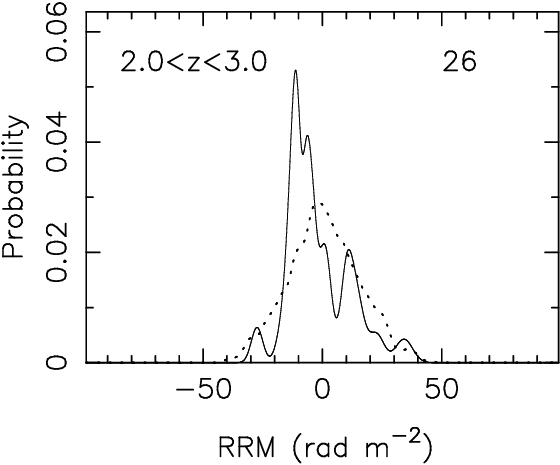}
\includegraphics[angle=-90,width=33mm]{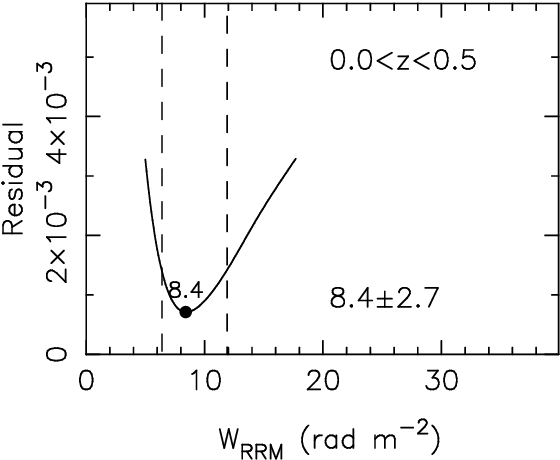}  
\includegraphics[angle=-90,width=33mm]{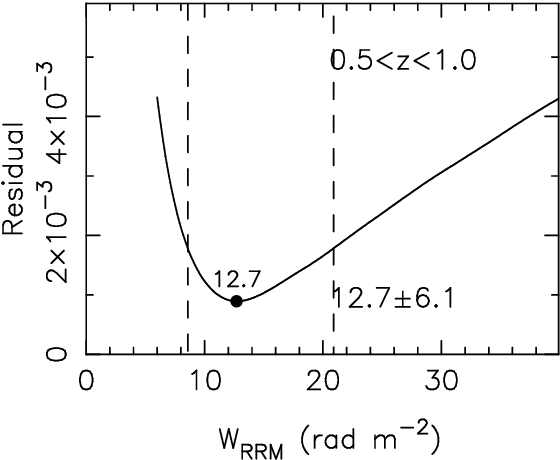}  
\includegraphics[angle=-90,width=33mm]{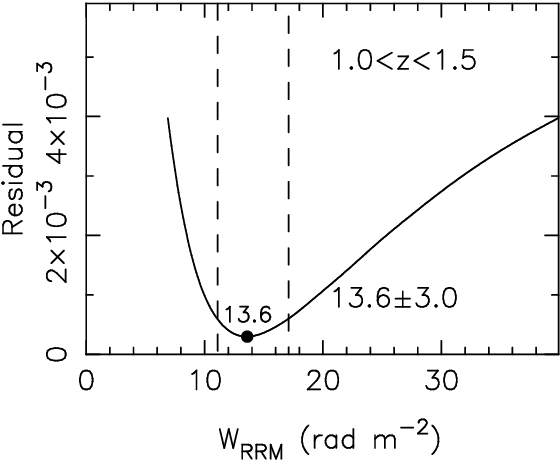}  
\includegraphics[angle=-90,width=33mm]{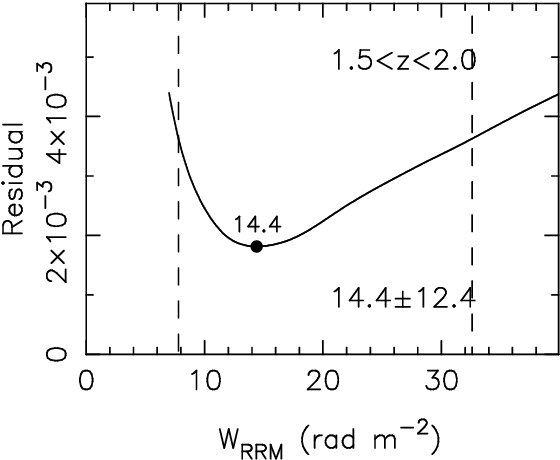}  
\includegraphics[angle=-90,width=33mm]{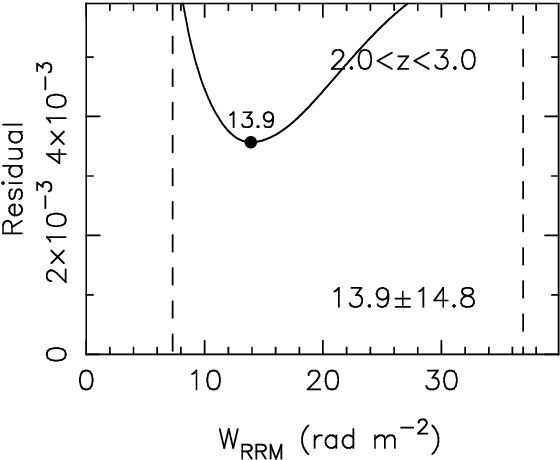} 
\caption{The same as Fig.~\ref{MockRRM20} but for the subsamples of
  quasars with the compiled RMs and $\sigma_{\rm RRM} \le $ 5
  rad~m$^{-2}$. Probability distribution function is not smooth due to
  small sample size and small RRM uncertainties.}
\label{MockRRM}
\end{figure*}

\begin{figure*}
\centering
\includegraphics[angle=-90,width=65mm]{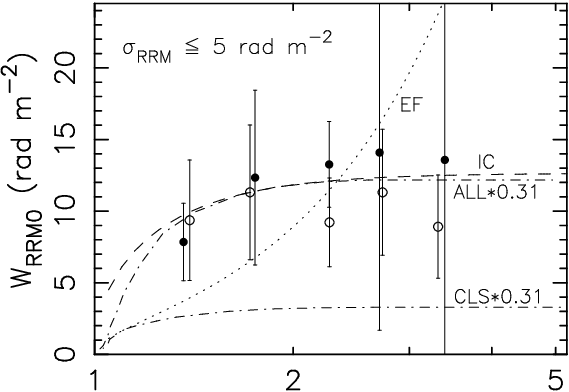} \hspace{1mm} 
\includegraphics[angle=-90,width=65mm]{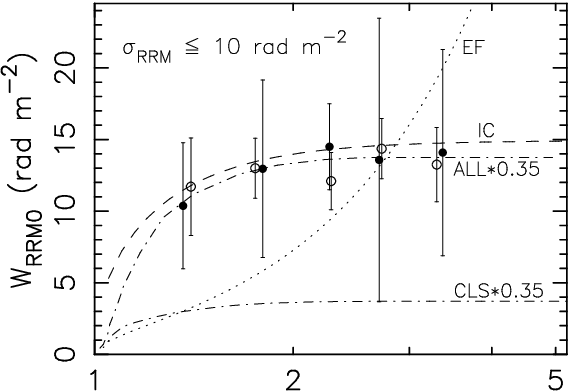} 
\includegraphics[angle=-90,width=65mm]{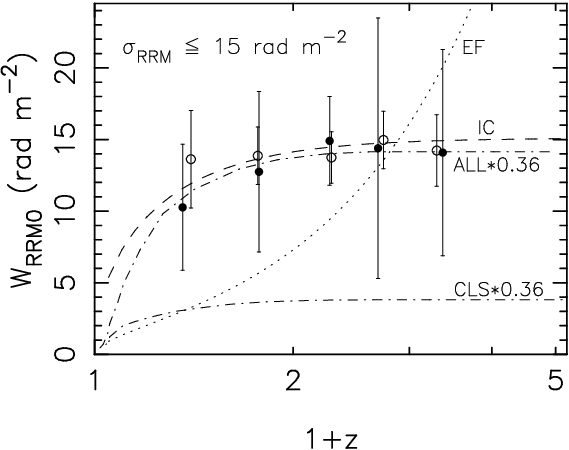} \hspace{1mm} 
\includegraphics[angle=-90,width=65mm]{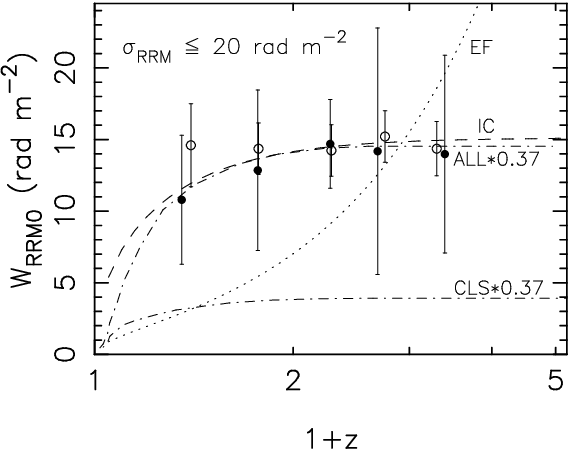} 
\caption{The real dispersion $W_{RRM0}$ of RRM distributions as a
  function of redshift for five subsamples of quasars in five redshift
  ranges, calculated for different RRM uncertainty thresholds and
  separately for the NVSS RMs (open circles) and the compiled RMs
  (filled circles). The median redshift of the subsample is adopted
  for each redshift bin. Two dot-dashed lines are the scaled ``ALL''
  and ``CLS'' model from \citet{ar11}, the dotted line is the evolving
  Friedmann model \citep[the EF model by][]{nel73}, and the dashed
  line is the ionized cloud model \citep[i.e. the IC model
    by][]{tn82}, which are scaled and fitted to the filled circles.}
\label{RRMzwidth}
\end{figure*}

\section{RRM distributions and their redshift evolution}

The RRM data shown in Figure~\ref{fig2} should be carefully analysed
to reveal the possible redshift evolution of the RRM distribution.

Looking at Figure~\ref{fig2}, we see that the most of 2642 RRMs have
values less than 50~rad~m$^{-2}$, with a peak around 0~rad~m$^{-2}$.
Only a small sample of quasars have $|RRM|>$ 50 rad~m$^{-2}$, which
may result from intrinsic RMs of sources or RM contribution from
galaxy clusters. The RM dispersion due to foreground galaxy clusters
is about 100 rad~m$^{-2}$ \citep[see][]{gdm+10,ckb01}. In this paper
we do not investigate the RRMs from galaxy clusters, therefore exclude
91 objects (3.44\%) with $|RRM|>$ 50 rad~m$^{-2}$, and then 2551
quasars are left in our sample for further analysis. Secondly, most of
these quasars have a redshift $z<3$. Because the sample size for high
redshift quasars is too small to get meaningful RRM statistics, we
excluded 62 quasars (2.43\%) of $z>3$ for further analysis of redshift
evolution. Finally we have RRMs of 2489 quasars with $|RRM| \le$ 50
rad~m$^{-2}$ and $z<3$.

Noticed in Table~\ref{qsosample} that RRMs of these quasars have
formal uncertainties $\sigma_{\rm RRM}$ between 0 and 20~rad~m$^{-2}$,
which would undoubtedly broaden the real RRM value distribution
and probably bury the possible small excess RRM with redshift. We
therefore work on 4 subsamples of these quasars with different RRM
uncertainty thresholds, $\sigma_{\rm RRM} \leqslant$ 20 rad~m$^{-2}$,
15 rad~m$^{-2}$, 10 rad~m$^{-2}$ and 5 rad~m$^{-2}$.
Because the NVSS RMs has an implicit systematic uncertainty of
$10.0\pm1.5$ rad m$^{-2}$ \citep{xh14}, different from that of the
compiled RMs which is less than 3 rad m$^{-2}$, we study the RRM
distribution for two samples of quasars separately: one with RMs taken
from the NVSS RM catalog, and the other with RMs from the compiled RM
catalog. We divide the quasar samples into five subsamples in five
redshift bins, $z=(0.0, 0.5)$, (0.5, 1.0), (1.0, 1.5), (1.5, 2.0), and
(2.0, 3.0), to check the redshift evolution of real dispersion
of RRM distributions.

How to get the real dispersion of RRM distributions, given
various uncertainties of RRM values? We here used the bootstrap
method. It is clear that the probability of a real RRM value follows a
Gaussian function centered at the observed RRM value with a width of
the uncertainty value, i.e.
\begin{equation}
p(RRM)=\frac{1}{\sqrt{2\pi}\sigma_{RRM_i}}e^{-\frac{(RRM-RRM_i)^2}{2\sigma_{RRM_i}^2}},
\label{normaldis}
\end{equation}
here $RRM_i$ is the $i$th data in the sample, and $\sigma_{RRM_i}$ is
its uncertainty. We then sum so-calculated probability distribution
function (i.e. the PDF in literature) for $N$ observed RRM values for
a subsample of quasars in a redshift range, assuming that there are
in-significant evolution in such a small redshift range,
\begin{equation}
P(RRM)=\sum_0^Np(RRM_i),
\label{sump}
\end{equation}
which contains the contributions from not only real RRM
distribution width but also the effect of observed RRM uncertainties.

If there is an ideal RRM data set without any measurement uncertainty,
the RRM values follow a Gaussian distribution with the zero mean and a
standard deviation of $W_{\rm RRM}$ which is the real dispersion
of RRM data due to medium between sources and us. We generate such a
mock sample of RRM data with the sample size 30 times of original RRM
data but with a RRM uncertainty randomly taken from the observed
RRMs. We sum the RRM probability distribution function for the mock
data, as done for real data. We finally can compare the two
probability distribution functions, $P(RRM)$ and $P_{\rm mock}(RRM)$,
by using the $\chi^2$ test as for two binned data sets \citep[see
  Sect.14.2 in ][]{ptvf92}. For each of input $W_{\rm RRM}$, the
comparison gives a residual $(P(RRM)-P_{\rm mock}(RRM))^2$ which
mimics the $\chi^2$. For a set of input values of $W_{\rm RRM}$, we
obtain the residual curve.  Example plots for the subsample of quasars
with the NVSS RMs and $\sigma_{RRM} \leqslant$ 20 rad~m$^{-2}$ are
shown in Figure~\ref{MockRRM20}, and for quasars with the compiled RMs
and $\sigma_{RRM} \leqslant$ 5 rad~m$^{-2}$ shown in
Figure~\ref{MockRRM}. Obviously the best match between $P(RRM)$ and
$P_{\rm mock}(RRM)$ with an input $W_{\rm RRM}$ should gives smallest
residual, so that we take this best $W_{\rm RRM}$ as the real RRM
dispersion. The residual curve, if normalized with the uncertainty of
the two PDFs which is unknown and difficult, should give the
$\chi^2=1$ for the best fit, and $\Delta\chi^2=1$ in the range for the
doubled residual for the 68\% confidence level. Therefore the
uncertainty of $W_{\rm RRM}$ is simply taken for the range with 
less than 2 times of the minimum residual in the residual curve.

In the last, note that there is an implicit systematic uncertainty of
$\sigma_{\rm sys} = 10.0 \pm 1.5$ rad~m$^{-2}$ in the NVSS RMs and the
maximum about $\sigma_{\rm sys} <$ 3 rad~m$^{-2}$ in the compiled RMs
\citep{xh14}, which are inherent in observed RRM values. The above
mock calculations have not considered this contribution, and therefore
the real dispersion of RRM distribution should be $W_{\rm RRM0}=
\sqrt{W_{\rm RRM} ^2-\sigma_{sys}^2}$. We listed all calculated
results of $W_{\rm RRM}$ and $W_{\rm RRM0}$ for all subsamples of
quasars in Table~\ref{dataresult}. Because almost all $W_{\rm RRM}$
and $W_{\rm RRM0}$ have a value larger than 10 rad~m$^{-2}$, the small
uncertainty of the systematic uncertainty of less than 2 or 3
rad~m$^{-2}$ does not make remarkable changes on these results in
Table~\ref{dataresult}.

Figure~\ref{RRMzwidth} plots different $W_{\rm RRM0}$ values as a
function of redshift ($1+z$) for five subsamples of quasars,
calculated for quasar subsamples with different thresholds of RRM
uncertainties and also separately for quasars with the NVSS RMs and
with the compiled RMs. We noticed that the $W_{\rm RRM0}$ values
obtained from the NVSS RMs and the compiled RMs are roughly consistent
within error-bars, and that the $W_{\rm RRM0}$ values obtained from
RRMs with different RRM thresholds are also consistent within
error-bars.  In all four cases of different $\sigma_{\rm RRM}$
thresholds, we can not see any redshift evolution of the $W_{\rm
  RRM0}$ of quasars with only the NVSS RMs, which is consistent with
the conclusions obtained by \citet{hrg12} and \citet{bml12}.  However,
the $W_{\rm RRM0}$ values systematically increase (from $\sim$10 to
$\sim$15 rad~m$^{-2}$) with the $\sigma_{RRM}$ thresholds (from 5 to
20 rad~m$^{-2}$), which implies the leakage of $\sigma_{\rm RRM}$ to
$W_{\rm RRM0}$ even after the simple discounting systematical
uncertainty. There is a clear tendency of the change of $W_{\rm RRM0}$
for quasars with the compiled RMs, increasing steeply when $z < 1$ and
flattening after $z > 1$, best seen from the samples of $\sigma_{RRM}
\leqslant$ 5 rad~m$^{-2}$. This indicates the marginal redshift
evolution, which is consistent with the conclusion given by,
e.g. \citet[][]{kbm+08} and \citet{jc13}. We therefore understand that
the small amplitude dispersion of RRMs is buried by the large
uncertainty of RRMs, and such real RRM evolution can only be
detected through high precision RM measurements of a large sample of
quasars in future.

\section{Discussions and conclusions}

Using the largest sample of quasar RMs and the best determined
foreground Galactic RMs and after carefully excluding the influence of
RRM uncertainties and large RRM ``outliers'', we obtained
Figure~\ref{RRMzwidth} to show the redshift evolution of dispersion of
extragalactic rotation measures. We now try to compare our results
with previously available models mentioned in Section 1.

As nowdays, the $\Lambda$CDM cosmology is widely accepted. The
non-evolving steady-state universe is no longer supported by so many
modern observations and we will not discuss it. The old coexpanding
evolving Friedmann model \citep{nel73} is ruled out by our RRM data as
well (see Figure \ref{RRMzwidth}), because the electron density and
magnetic field in the model are scaled with redshift via
$n_e=n_0(1+z)^3$ and $B=B_0(1+z)^2$ and the variance of RRMs ($\propto
W_{\rm RRM0}^2$) should increase with $z$. Among the three old models,
the ionized cloud (IC) model given by \citet{tn82} can really include
all possible RM contributions and fit to the data. The ionized clouds
along the line of sight can be the gravitationally bounded and ionized
objects, which may be associated with protogalaxies, galactic halos,
galaxy clusters or even widely distributed intergalactic medium in
cosmic webs. The dashed lines in Figure \ref{RRMzwidth} are the
fitting to the $W_{\rm RRM0}$ data by the ionized cloud model. In
$\Lambda$CDM cosmology, it has the form of
\begin{equation}
V_{\rm RRM}=V_0 \int_0^{z_s}\frac{1}{(1+z)^3\sqrt{\Omega_m(1+z)^3+\Omega_{\Lambda}}}dz
\label{ICform}
\end{equation}
 with a fitting parameter
\begin{equation}
V_0 = (0.81 n_c B_{||c})^2 \frac{c \; l_c\; f_0}{H_0} \approx 441\pm150~{\rm rad^2 m^{-4}}, 
\label{IC}
\end{equation}
where $n_c$, $B_c$ and $l_c$ are the electron density, magnetic field
and the coherence size of a random field size, $f_0$ is the filling
factor, $H_0$ is the Hubble parameter and $c$ is the light velocity.
Current $\Lambda$CDM cosmology takes $H_0$=70 km~s$^{-1}$~Mpc$^{-1}$,
$\Omega_m$=0.3 and $\Omega_{\Lambda}$=0.7. The RRM variance ($V_{\rm
  RRM} \propto W_{\rm RRM0}^2$) in the ionized cloud model has a steep
increase at low redshift and flattens at high redshift, which fits the
$W_{\rm RRM0}$ data very well (see Figure~\ref{RRMzwidth}). The
similations given by \citet{ar11} verified the shape of the RM
dispersion curves. We scaled the ``ALL'' model of \citet{ar11} to fit
the data, and also scaled their ``CLS'' model to show the relatively
small amplitude from cosmic webs.

For a sample of quasars, the lines of sight for some of them pass
through galaxy halos indicated by MgII absorption lines which probably
have a RRM dispersion of several rad~m$^{-2}$ \citep{jc13}; some
quasars behind galaxy clusters may have large RRM dispersion of a few
tens rad~m$^{-2}$ \citep{ckb01,gdm+10}; some quasars just through
intergalactic medium without such intervening objects should have a
RRM dispersion of 2$\sim$3 rad~m$^{-2}$ from the cosmic webs (see the
cluster subtracted model of \citet{ar11}. These different clouds give
different $V_0$. We noticed, however, that the redshift evolution of
RRM dispersions of each kind of clouds depends only on cosmology (see
Eq.~\ref{ICform}), not the $V_0$.

In principle, we can model the RRM dispersion with a combination of
ionized clouds with different fractions, i.e. $V_0=V_{\rm gala}*f_{\rm
  gala}+V_{\rm cluster}*f_{\rm cluster}+V_{\rm IGM}*f_{\rm IGM}$. We
checked our quasar samples in the SDSS suvery area, about 10\% to 15\%
of quasars (for different samples in Table 2) are behind the known
galaxy clusters of $z \le 0.5$ in the largest cluster catalog
\citep{whl12}. Quasars behind galaxy clusters have a large scatter in
RRM data in Figure \ref{fig2}, mostly probably extended to beyound 50
rad~m$^{-2}$, which give a wide Gaussian distribution of real RRM
dispersions. The fraction for the cluster contribition is at least
$f_{\rm cluster}\sim 0.1-0.15$, because of unknown clusters at higher
redshifts. The fraction for galaxy halo contribution shown by MgII
absorption lines $f_{\rm gala}$ is about 28\% \citep{jc13}. If we {\it
  assume} the coherence size of magnetic fields in these three clouds
as 1 kpc, 10 kpc and 1000 kpc, the mean electron density as $10^{-3}$
cm$^{-3}$, $10^{-4}$ cm$^{-3}$ and $10^{-5}$ cm$^{-3}$, and mean
magnetic field as 2 $\mu$G, 1$\mu$G and 0.02 $\mu$G
\citep[e.g.][]{ar11}, and the filling factors as 0.00001, 0.001 and
0.1 \citep{tn82} for galaxy halos, galaxy clusters and intergalactic
medium in cosmic webs, we then can estimate the dispersions of these
clouds, which are 7, 11 and 2 rad~m$^{-2}$ at z=1, respectively.
Whatever values for the different ionized clouds, they will have to
sum together with various fractions to fit the dispersions of RRM
data.

\begin{figure}
\centering
\includegraphics[angle=-90,width=65mm]{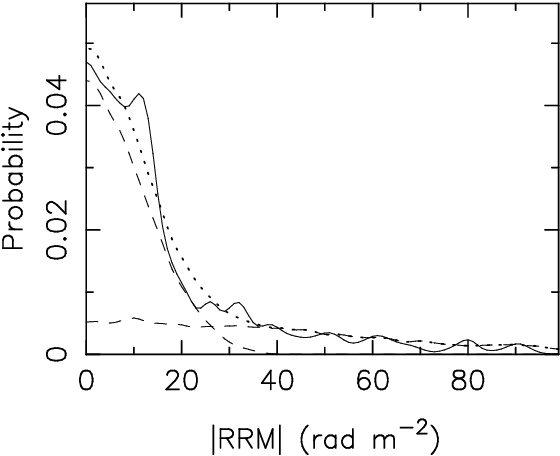} 
\includegraphics[angle=-90,width=65mm]{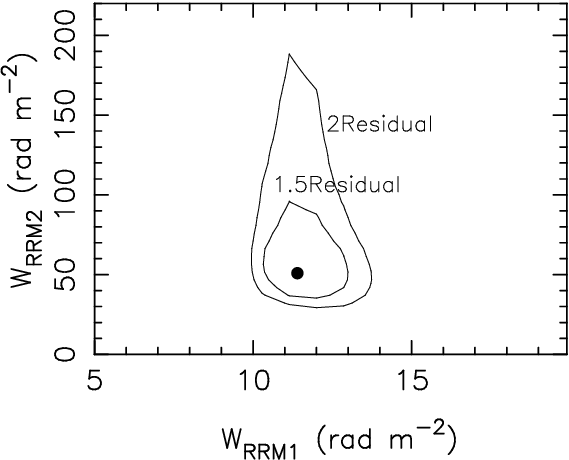} 
\caption{The probability distribution function for the absolute values
  of RRMs in the {\it upper panel} for 146 quasars of $z \ge 1$ from
  the compiled RM catalog \citep{xh14} with RM uncertainty
  $\sigma_{RRM} \leqslant$ 5 rad~m$^{-2}$, which is fitted by the two
  mock samples with a narrow real RRM dispersion $W_{\rm RRM1}$
  standing for the contributions from galaxy halos and cosmic webs and
  a wide RRM dispersion $W_{\rm RRM2}$ for the contribution from
  galaxy clusters.  The likelihood contours for the best fits by using
  two dispersions are shown in the {\it lower panel}, with the best
  $W_{RRM}$ values marked as the black dot.}
\label{components}
\end{figure}

After realizing that the real RRM dispersion of quasars at $z>1$ does
not change with redshift for each kind of ionized clouds, we now model
the probability distribution function of absolute values of RRM data
for all 146 quasars with $z \ge 1$ from the compiled RM catalog,
without discarding any objects limited by redshift and RRM values but
with a formal RM uncertainty $\sigma_{RRM} \leqslant$ 5 rad~m$^{-2}$
(see Figure~\ref{components}). We found that such a probability
function can be fitted with two components, one for a small $W_{\rm
  RRM}$ which stands for the contributions from galactic halos and
cosmic webs, and one for a wide $W_{\rm RRM}$ which comes from the
galaxy clusters. Two such muck samples with optimal fractions are
searched for the best match of the probability function. We get
$W_{RRM1} =11.4^{+3.3}_{-1.4} $rad~m$^{-2}$ with a fraction of
$f_1$=0.65, and $W_{RRM2}=51^{+135}_{-20}$ rad~m$^{-2}$ with a
fraction of $f_2$=0.35 for clusters. However, we can not separate the
contributions from galaxy halos and cosmic webs which are tangled
together in $W_{RRM1}$.

We therefore conclude that the dispersion of RRM data steady increases
and get the saturation at about 10 rad~m$^{-2}$ when $z \ge
1$. However, the current RM dataset, even the largest sample of
quasars, are not yet good enough to separate the RM contributions from
galaxy halos and cosmic webs due to large RRM uncertainties. A
larger sample of quasars with better precision of RM measurements are
desired to make clarifications.

\section*{Acknowledgments}

The authors thank Dr. Hui Shi for helpful discussions. The authors are
supported by the National Natural Science Foundation of China
(10833003) and by the Strategic Priority Research Program ``The
Emergence of Cosmological Structures'' of the Chinese Academy of
Sciences, Grant No. XDB09010200”。
%
This research has made use of the NASA/IPAC Extragalactic Database (NED) 
which is operated by the Jet Propulsion Laboratory,California Institute 
of Technology, under contract with the National Aeronautics and Space 
Administration.
Funding for SDSS-III has been provided by the Alfred P. Sloan Foundation, the 
Participating Institutions, the National Science Foundation, and the U.S. 
Department of Energy Office of Science. The SDSS-III web site is 
http://www.sdss3.org/ .

\bibliographystyle{mn2e}
\bibliography{qsoref}

\label{lastpage}
\end{document}